\documentclass{emulateapj}


\usepackage{epsfig}
\usepackage{amsmath}
\usepackage{xspace}
\usepackage{graphics}
\usepackage{natbib}
\usepackage{bm}
\usepackage{color}
\usepackage[normalem]{ulem}
\newcommand{\ignore}[1]{}
\newcommand{\kagn}{KA1858+4850\xspace}
\newcommand{\perpix}{pixel\ensuremath{^{-1}}\xspace}
\newcommand{\halpha}{H\ensuremath{\alpha}\xspace}
\newcommand{\hbeta}{\ensuremath{\mathrm{H}\beta}\xspace}
\newcommand{\hgamma}{H\ensuremath{\gamma}\xspace}
\newcommand{\hdelta}{H\ensuremath{\delta}\xspace}
\newcommand{\heii}{\ion{He}{2}\xspace}
\newcommand{\kms}{km s\ensuremath{^{-1}}\xspace}
\newcommand{\msun}{\ensuremath{{M}_{\odot}}\xspace}
\newcommand{\mbh}{\ensuremath{M_{\mathrm{BH}}}\xspace}
\newcommand{\angstrom}{\AA}
\newcommand{\taupeak}{\ensuremath{\tau_{\mathrm{peak}}}\xspace}
\newcommand{\taucen}{\ensuremath{\tau_{\mathrm{cen}}}\xspace}
\newcommand{\taujav}{\ensuremath{\tau_{\mathrm{\texttt{JAVELIN}}}}\xspace}
\newcommand{\fluxunit}{\ensuremath{10^{-15}~\mathrm{erg~cm}^{-2}~\mathrm{s}^{-1}}\xspace}
\newcommand{\msigma}{\ensuremath{M_{\mathrm{BH}}-\sigma_{\star}}\xspace}
\newcommand{\lag}{\ensuremath{14.58^{+2.19}_{-2.50}}\xspace} %observed frame CCF
\newcommand{\lagrest}{\ensuremath{13.53^{+2.03}_{-2.32}}\xspace} %rest frame CCF
\newcommand{\lagjavelinrest}{\ensuremath{13.15^{+1.08}_{-1.00}}\xspace} %rest frame JAVELIN lag
\newcommand{\bhmass}{\ensuremath{M_{\mathrm{BH}} = 8.06^{+1.59}_{-1.72} \times 10^6~\msun}\xspace}
\newcommand{\bhmassjavelin}{\ensuremath{M_{\mathrm{BH,\texttt{JAVELIN}}} = 6.58^{+1.00}_{-0.98} \times 10^6~\msun}\xspace}
\newcommand{\hbetawidth}{\ensuremath{770 \pm 49}~\kms} %subtracted intrinsic line width

\begin{document}

%Title and a list of authors with superscript footnotes for affiliations
\title{Reverberation Mapping of the \textit{Kepler}-Field AGN \kagn}
\author{
  Liuyi Pei\altaffilmark{1}, %UCI
  Aaron J. Barth\altaffilmark{1}, %UCI
  Greg S. Aldering\altaffilmark{2}, %LBNL
  Michael M. Briley\altaffilmark{3}, %Appalachian
  Carla J. Carroll\altaffilmark{4}, %BYU
  Daniel J. Carson\altaffilmark{1}, %UCI
  S. Bradley Cenko\altaffilmark{5,6,7}, %Maryland,NASA,SSI
  Kelsey I. Clubb\altaffilmark{8}, %UC Berkeley
  Daniel P. Cohen\altaffilmark{8}, %UC Berkeley
  Antonino Cucchiara\altaffilmark{2,8}, %LBNL/UC Berkeley
  Tyler D. Desjardins\altaffilmark{9}, %University of Western Ontario
  Rick Edelson\altaffilmark{5}, %Maryland
  Jerome J. Fang\altaffilmark{10}, %UC Santa Cruz
  Joseph M. Fedrow\altaffilmark{11,12}, %SDSU, CASS
  Alexei V. Filippenko\altaffilmark{8}, %UC Berkeley
  Ori D. Fox\altaffilmark{8}, %UC Berkeley
  Amy Furniss\altaffilmark{13}, %Stanford
  Elinor L. Gates\altaffilmark{14}, %Lick
%  Melissa L. Graham\altaffilmark{10}, %LCOGT; does not want to be co-author
  Michael Gregg\altaffilmark{15}, %UC Davis
  Scott Gustafson\altaffilmark{16}, %UCSD
  J. Chuck Horst\altaffilmark{11}, %SDSU
  Michael D. Joner\altaffilmark{4}, %BYU
  Patrick L. Kelly\altaffilmark{8}, %Berkeley  Delete 17, 18
  Mark Lacy\altaffilmark{17}, %NRAO
  C. David Laney\altaffilmark{4}, %BYU
  Douglas C. Leonard\altaffilmark{11}, %SDSU
%  David Levitan\altaffilmark{17}, %Caltech
  Weidong Li\altaffilmark{8,18}, %UC Berkeley
  Matthew A. Malkan\altaffilmark{19}, %UCLA
  Bruce Margon\altaffilmark{10}, %UCSC
  Marcel Neeleman\altaffilmark{16}, %UCSD
  My L. Nguyen\altaffilmark{20}, %University of Wyoming
  J. Xavier Prochaska\altaffilmark{10,21}, %UC Santa Cruz/ UCO/Lick
  Nathaniel R. Ross\altaffilmark{18}, %UCLA
  David J. Sand\altaffilmark{22,23}, %LCOGT, TexasTech
  Kinchen J. Searcy\altaffilmark{24},%<ksearcy1@san.rr.com>
  Isaac S. Shivvers\altaffilmark{8}, %Berkeley
  Jeffrey M. Silverman\altaffilmark{25}, %UT Austin
  Graeme H. Smith\altaffilmark{21}, %UCO/Lick
  Nao Suzuki\altaffilmark{26}, %Kavli
  Krista Lynne Smith\altaffilmark{5}, %Maryland
  David Tytler\altaffilmark{16}, %UC San Diego
  Jessica K. Werk\altaffilmark{10,21}, %UCSC, UCO/Lick
  and G\'{a}bor Worseck\altaffilmark{21} %UCO/Lick
}

%Author affiliations go in a footnote 
\altaffiltext{1}{\ignore{UCI}Department of Physics and Astronomy, University of California, Irvine, CA, 92697, USA}
\altaffiltext{2}{\ignore{LBNL}Lawrence Berkeley National Laboratory, Berkeley, CA 94720, USA}
\altaffiltext{3}{\ignore{Appalachian}Department of Physics and Astronomy, Appalachian State University, Boone, NC 28608, USA}
\altaffiltext{4}{\ignore{BYU}Department of Physics and Astronomy, N283 ESC, Brigham Young University, Provo, UT 84602, USA}
\altaffiltext{5}{\ignore{Maryland}Department of Astronomy, University of Maryland, College Park, MD 20742, USA}
\altaffiltext{6}{\ignore{NASA}Astrophysics Science Division, NASA Goddard Space Flight Center, Mail Code 661, Greenbelt, MD 20771, USA}
\altaffiltext{7}{\ignore{SSI}Joint Space Science Institute, University of Maryland, College Park, MD 20742, USA}
\altaffiltext{8}{\ignore{UCB}Department of Astronomy, University of California, Berkeley, CA 94720-3411, USA}
\altaffiltext{9}{\ignore{UWO}Department of Physics and Astronomy, The University of Western Ontario, London, ON, N6A 3K7, Canada}
\altaffiltext{10}{\ignore{UCSC}Department of Astronomy and Astrophysics, University of California, Santa Cruz, 1156 High Street, Santa Cruz, CA 95064, USA}
\altaffiltext{11}{\ignore{SDSU}Department of Astronomy, San Diego State University, San Diego, CA 92182, USA}
\altaffiltext{12}{\ignore{CASS}Center for Astrophysics and Space Sciences, University of California, San Diego, La Jolla, CA 92093, USA}
\altaffiltext{13}{\ignore{Stanford}Department of Physics, Stanford University, 382 Via Pueblo Mall, Stanford, CA 94305, USA}
\altaffiltext{14}{\ignore{Lick}Lick Observatory, P.O. Box 85, Mt. Hamilton, CA 95140}
\altaffiltext{15}{\ignore{UCDavis}Department of Physics, University of California, Davis, One Shields Ave, Davis, CA 95616, USA}
\altaffiltext{16}{\ignore{UCSD}Department of Physics, University of California, San Diego, La Jolla, CA 92093, USA}
\altaffiltext{17}{\ignore{NRAO}NRAO, 520 Edgemont Road, Charlottesville, VA 22903, USA}
\altaffiltext{18}{Deceased 12 December 2011}
\altaffiltext{19}{\ignore{UCLA}Department of Physics and Astronomy, University of California, Los Angeles, CA 90095, USA}
\altaffiltext{20}{\ignore{Wyoming}Department of Physics and Astronomy, University of Wyoming, 1000 E. University Ave. Laramie, WY 82071, USA}
\altaffiltext{21}{\ignore{UCO/Lick}Department of Astronomy and Astrophysics, UCO/Lick Observatory, University of California, 1156 High Street, Santa Cruz, CA 95064, USA}
\altaffiltext{22}{\ignore{LCOGT}Las Cumbres Observatory Global Telescope Network, 6740 Cortona Drive, Suite 102, Goleta, CA 93117, USA}
\altaffiltext{23}{\ignore{TexasTech}Department of Physics, Texas Tech University, Lubbock, TX 79409, USA}
\altaffiltext{24}{\ignore{SDAA}San Diego Astronomy Association, P.O Box 23215, San Diego, CA 92193, USA}
\altaffiltext{25}{\ignore{UTAustin}Department of Astronomy, University of Texas at Austin, 2515 Speedway, Stop C1400, Austin, Texas 78712, USA}
\altaffiltext{26}{\ignore{Kavli}Kavli Institute for Particle Astrophysics and Cosmology, Stanford University, 452 Lomita Mall, Stanford, CA 94305, USA}

\begin{abstract}

\kagn is a narrow-line Seyfert 1 galaxy at redshift 0.078 and is among the brightest active galaxies monitored by the \textit{Kepler} mission. We have carried out a reverberation mapping campaign designed to measure the broad-line region size and estimate the mass of the black hole in this galaxy. We obtained 74 epochs of spectroscopic data using the Kast Spectrograph at the Lick 3-m telescope from February to November of 2012, and obtained complementary \textit{V}-band images from five other ground-based telescopes. We measured the \hbeta light curve lag with respect to the \textit{V}-band continuum light curve using both cross-correlation techniques (CCF) and continuum light curve variability modeling with the \texttt{JAVELIN} method, and found rest-frame lags of $\tau_\mathrm{CCF} = \lagrest$ days and $\tau_\mathrm{\texttt{JAVELIN}} = \lagjavelinrest$ days. The \hbeta root-mean-square line profile has a width of $\sigma_\mathrm{line} =$ \hbetawidth. Combining these two results and assuming a virial scale factor of $f=5.13$, we obtained a virial estimate of \bhmass for the mass of the central black hole and an Eddington ratio of $L/L_\mathrm{Edd} \approx 0.2$. We also obtained consistent but slightly shorter emission-line lags with respect to the \textit{Kepler} light curve. Thanks to the \textit{Kepler} mission, the light curve of \kagn has among the highest cadences and signal-to-noise ratios ever measured for an active galactic nucleus; thus, our black hole mass measurement will serve as a reference point for relations between black hole mass and continuum variability characteristics in active galactic nuclei.

\end{abstract}

%====================================================================================================

\section{Introduction}

The NASA \textit{Kepler} Mission, designed to search for exo-planets, continuously monitored the brightness of more than $100,000$ stars in a 115 square-degree field for about four years \citep{borucki2010}. Situated within the \textit{Kepler} field are several active galactic nuclei (AGNs) that also exhibit optical flux variations. \textit{Kepler}'s monitoring capabilities enable measurements of AGN optical light curves over long temporal baselines with unprecedented cadence and precision, providing the basis for extremely detailed AGN variability studies.

Observations have revealed correlations between AGN variability amplitude and redshift \citep{cristiani1990,giallongo1991,hook1994,fernandes1996,vandenberk2004}, variability amplitude and black hole mass \citep{wold2007,wilhite2008,bauer2009}, and anticorrelations between variability amplitude and luminosity \citep{cristiani1990,hook1994,fernandes1996,giveon1999,vandenberk2004,webbmalkan2000}. Furthermore, analyses of continuum light curves have revealed the presence of characteristic variability timescales which have been found to vary with black hole mass \citep{collierpeterson2001,macleod2010}. The \textit{Kepler} high-resolution light curves have a cadence of 30 minutes, and are the only datasets to date that have been able to probe optical AGN variability down to such short time scales. Optical fluctuation power spectral density functions for several \textit{Kepler} AGNs have already been published \citep{mushotzky2011}, and they have shown much steeper slopes than those seen in the X-rays. \textit{Kepler}'s light curves provide new high signal-to-noise ratio (SNR) data which will test and better constrain these previously established correlations and further shed light on AGN variability characteristics. 

Independent measurements of black hole mass are required to search for connections between AGN variability characteristics and black hole mass. To this end, we present the results of a nine-month monitoring campaign for the narrow-line Seyfert 1 (NLS1) galaxy 1RXSJ185800.9+485020, also known as \kagn, which has redshift $z=0.078$ and a Galactic extinction of $A_V = 0.15$ mag \citep{schlafly2011}. This object was identified as an X-ray source in the \textit{ROSAT} All-Sky Bright Source Catalogue \citep{rosat}. Prior to 2012, there was no published spectrum of \kagn in the literature, and an observation from Lick Observatory identified it as a Seyfert 1 galaxy \citep{edelsonmalkan2012}. The initial portion of \kagn's \textit{Kepler} light curve from quarters Q6 and Q7 was published by \citet{mushotzky2011} and showed strong optical variability, qualifying it as a prime candidate for reverberation mapping.

The technique of reverberation mapping relies on the assumption that variability in the AGN continuum is echoed by emission lines originating from the surrounding broad-line region \citep[BLR;][]{blandfordmckee1982}. Ionizing photons from the AGN central engine travel to the BLR gas in a time $\tau$ that is a function of the BLR radius. Changes in the ionizing photon flux incident on BLR clouds cause fluctuations in the emission-line flux. This means that the emission-line light curve will appear as a lagged version of the continuum light curve, and the lag time, combined with the speed of light, can give an estimate of the BLR radius. Additionally, the line-emitting gas orbits the central black hole at very high velocities, which causes Doppler broadening of the emitted spectral lines. The width of the broad emission line gives the velocity dispersion of the BLR gas, which, combined with the BLR radius, can yield a virial estimate of the central black hole mass.

\textit{Kepler} light curves covering over two years of monitoring are now publicly available for \kagn, of which three consecutive quarters (Q13, Q14, and Q15) directly coincide with the time of our ground-based monitoring campaign. We therefore performed our analysis using both \textit{V}-band and \textit{Kepler} observations.

We employed the Lick Observatory 3-m Shane telescope with the Kast Spectrograph and five other ground-based telescopes to spectroscopically and photometrically monitor \kagn from February to November of 2012. We describe our imaging observations and data reductions in \S2 and \S3; spectroscopic observations, reductions, and measurements are described in \S4 and \S5; \S6 outlines the steps in measuring emission-line light curve lags; our estimates of the black hole mass (\mbh) and Eddington ratio are discussed in \S7 and \S8; and \S9 summarizes our results.

%====================================================================================================

\section{Imaging Observations}

Reverberation mapping requires a continuum light curve with high sampling cadence and SNR. To achieve this, we obtained \textit{V}-band images from ground-based telescopes and used aperture photometry to construct a light curve for \kagn that has nearly nightly sampling for a span of 290 days. For several reasons, we chose to use the \textit{V}-band light curve rather than the \textit{Kepler} light curve for reverberation measurements. First, we wanted to monitor the AGN's variability in real time, and since \textit{Kepler} data are uploaded only periodically, this was possible only with ground-based monitoring. Additionally, the \textit{Kepler} passband, at 4000$-$8650 \angstrom, includes the strong \halpha emission line, which can contribute significantly to the photometric fluxes and introduce a strong lag signal to what should ideally be a pure continuum light curve. Furthermore, \textit{Kepler} light curves exhibit severe mismatches between the flux scales for different quarterly observation sets, as can be seen in light curves shown by \citet{revalski2014}. We avoided these issues by constructing the continuum light curve with photometric data from five ground-based telescopes, whose properties are described in the following sections.

%------------------------------------------------------------------------------------------

\subsection{West Mountain Observatory}
The Brigham Young University West Mountain Observatory (WMO) uses a 0.9-m telescope that employs a FLI PL3041UV detector with a $20\farcm8 \times 20\farcm8$ field of view. The CCD has 15 $\mu$m pixels and a scale of $0\farcs61$ \perpix. \kagn was observed at WMO with exposure times of 200 s, 240 s, 250 s, or 300 s. WMO data covered the period from March to November of 2012 with images from 124 nights, and had a median seeing of $3\farcs2$. Figure~\ref{fig:fieldstars} shows a portion of the WMO field of view centered on \kagn.

%------------------------------------------------------------------------------------------

\subsection{KAIT}
The Katzman Automatic Imaging Telescope (KAIT) at Lick Observatory is a 0.76-m robotic telescope with an Apogee AP7 CCD, which has 24 $\mu$m pixels in a $500 \times 500$ array and a scale of $0\farcs8$ \perpix \citep{filippenko2001}. \kagn was observed with KAIT using 300 s exposures with the exception of six nights, for which exposure times of 60 s, 180 s, or 240 s were used. The median seeing for the KAIT exposures was $3\farcs2$, and the observing period at KAIT spanned February to September of 2012 with data from 109 nights.
 
%------------------------------------------------------------------------------------------

\subsection{Faulkes Telescope North}
The Faulkes Telescope North (FTN), operated by the Las Cumbres Observatory Global Telescope Network, is a 2-m telescope located at the Haleakala Observatory in Hawaii. We used the Spectral camera with a Fairchild Imaging CCD486 detector, which has a $10\farcm5 \times 10\farcm5$ field of view \citep{brown2013}. The CCD has 15~$\mu$m pixels in a $4000 \times 4000$ array and has a scale of $0\farcs152$ \perpix. The images were obtained using $2 \times 2$ binning for the readout. \kagn was observed at FTN with 120 s exposures from February to March 2012. The exposure time was increased to 180 s in April 2012, then to 240 s in May for the remainder of the program ending in November. We obtained 65 epochs of data from FTN, with median seeing of $1\farcs6$.

%------------------------------------------------------------------------------------------

\subsection{The Nickel Telescope}
The 1-m Nickel telescope at Lick Observatory employs a Loral $2048 \times 2048$ CCD with a $6\farcm3 \times 6\farcm3$ field of view and a scale of $0\farcs184$ \perpix. The images were obtained using $2 \times 2$ binning for the readout. \kagn was observed on the Nickel adopting 300 s exposures with the exception of three nights, for which 150 s, 250 s, and 600 s exposures were used. We obtained 47 epochs of data from the Nickel between February and November of 2012, and the median seeing was $2\farcs4$.

%------------------------------------------------------------------------------------------

\subsection{Mount Laguna Observatory}
The Mount Laguna Observatory (MLO) 1-m telescope uses a Fairchild CCD that has 15 $\mu$m pixels in a $2048 \times 2048$ array, and has a scale of $0\farcs41$ \perpix. \kagn was observed at MLO with 300 s exposures. The median seeing for the MLO exposures was $3\farcs0$. Between February and November of 2012, we obtained 27 epochs of data from MLO.

%====================================================================================================

\section{Photometric Reductions and Measurements}

\subsection{\textit{V}-Band Data}

\begin{figure}
\epsfxsize=8.5cm\epsfbox{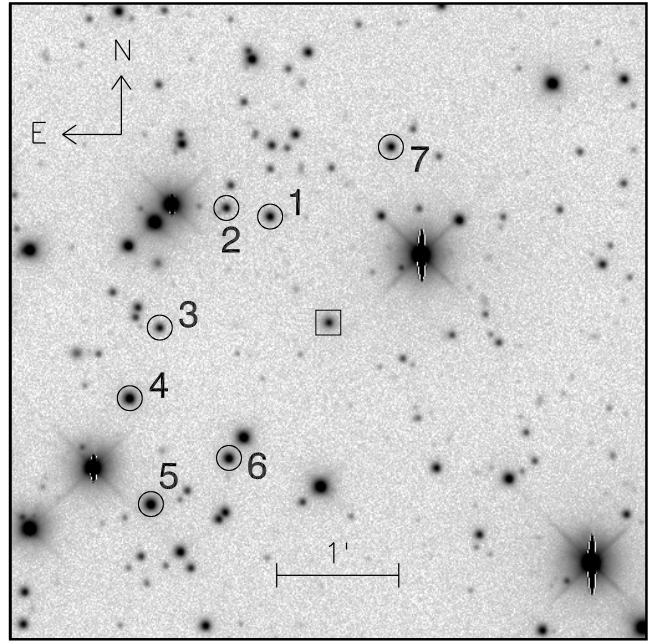}
\caption{A subset of a coadded frame created from WMO images showing \kagn (boxed) and its seven comparison stars (circled).}
\label{fig:fieldstars}
\end{figure}

Photometric data reduction included overscan correction, trimming, bias subtraction, and flat fielding for all images. We used the \textit{Astrometry.net} software \citep{lang2010} to register celestial coordinates onto images from WMO, KAIT, Nickel, and MLO. This step was omitted for FTN data, which already contained celestial coordinates in the image headers. After cleaning all images of cosmic rays using the L.A.Cosmic algorithm \citep{vandok2001}, we performed aperture photometry in IDL using an aperture radius of $3\arcsec$ and sky annulus radii of $10\arcsec$$-$$20\arcsec$, and obtained instrumental magnitudes for \kagn and seven comparison stars (marked in Figure~\ref{fig:fieldstars}) for each image. The comparison stars were chosen to have similar or slightly brighter \textit{V}-band magnitudes compared to \kagn. For nights where multiple exposures were taken at the same telescope, the magnitude measurements for each object were averaged into a single value. Since \kagn is almost indistinguishable from a point source at ground-based resolution, we did not attempt to remove host-galaxy light from the AGN photometry.

%-----FIGURE-----%

We used the comparison stars as constant-flux references and obtained a separate AGN light curve for each telescope. However, the uncertainties from aperture photometry photon counting errors underestimate the true photometric error budget. Additional sources of error include inconsistencies in flat-field corrections and poor comparison-star magnitude measurements owing to blemishes on the detector. We measured the magnitude of these additional errors by calculating the excess variance, defined as

\begin{equation}
\sigma^{2}_\mathrm{x} = \frac{1}{N} \displaystyle\sum\limits_{i=1}^{N} [(X_{i} - \mu)^{2} - \sigma^{2}_{i}],
\label{eq:excessvariance}
\end{equation}

\noindent in the scaled comparison-star light curves. Here, $N$ is the number of measurements in the sample, $\mu$ is the mean magnitude, and $X_i$ and $\sigma_i$ are the individual measurements and their associated uncertainties, respectively. The $\sigma_i$ values range from 0.004 mag to 0.048 mag, and the median and standard deviation of the uncertainties are 0.009 mag and 0.007 mag, respectively. We found the mean scatter of all seven comparison stars to be $\sigma_{\mathrm{x}} \approx 0.001$ mag, and added this in quadrature to the uncertainties from aperture photometry to produce the final AGN light curve for each telescope.

To combine the light curves from different telescopes, we scaled each light curve so that the mean comparison-star magnitudes for each telescope matched those from WMO, the telescope with the highest SNR and cadence and longest temporal coverage. However, each telescope has a different wavelength-dependent throughput, which can cause systematic offsets between light curves from different telescopes since the AGN is likely bluer than the average comparison-star color. We tested for these offsets by calculating the differences between AGN magnitude measurements taken on the same night but at different sites, and found the offsets to be on the order of $0.01$ mag. We applied these calculated shifts to the FTN, KAIT, MLO, and Nickel light curves and brought them into agreement with WMO to produce the combined light curve.

Finally, we used \citet{landolt1992} standard stars observed at WMO to calibrate the zero point of the magnitude scale and produce the final light curve. We used WMO images from 18 nights, on which the observers deemed conditions photometric, to calibrate the comparison-star magnitudes. We did not attempt to compute color dependence in the Landolt calibrations.

Because truly photometric conditions are rare and difficult to confirm, each night gave slightly different comparison-star magnitudes. We took the weighted mean magnitude and standard deviation over 18 nights to be the magnitude and uncertainty for each star. The \textit{V} magnitudes of the comparison stars are listed in Table~\ref{tab:compstarstable}.

\begin{figure}
\epsfxsize=8.9cm\epsfbox{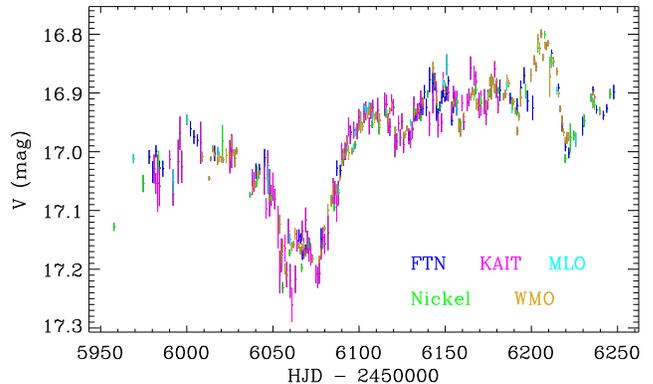}
\caption{\kagn \textit{V}-band light curve. The vertical length at each epoch indicates the photometric uncertainties.}
\label{fig:vbandltcurve}
\end{figure}

Figure~\ref{fig:vbandltcurve} plots the final \textit{V}-band light curve for \kagn. The vertical length at each epoch indicates the photometric uncertainties, and the data are listed in Table~\ref{tab:vbandtable}. We averaged photometric measurements taken within 12 hours of each other to produce a condensed light curve that was used for subsequent lag analyses.

%-----TABLE-----%

\begin{deluxetable}{clll}
\tablewidth{0pt}
\tablecolumns{4}
\tablecaption{Photometric Comparison Stars for \kagn \label{tab:compstarstable}}
\tablehead{
  \colhead{Star} &
  \colhead{$\alpha$} &
  \colhead{$\delta$} &
  \colhead{$V$} \\
  \colhead{} &
  \colhead{($h$:$m$:$s$)} &
  \colhead{($\circ$:$\prime$:$\prime\prime$)} &
  \colhead{(mag)}
}
\startdata
  1   &  18:58:04.03    &   48:51:15.53     &  16.612 $\pm$  0.026 \\
  2   &  18:58:06.24    &   48:51:19.82     &  17.121 $\pm$  0.037 \\
  3   &  18:58:09.54    &   48:50:20.69     &  17.256 $\pm$  0.031 \\
  4   &  18:58:11.03    &   48:49:46.16     &  15.449 $\pm$  0.028 \\
  5   &  18:58:09.97    &   48:48:54.06     &  15.195 $\pm$  0.029 \\
  6   &  18:58:06.09    &   48:49:16.65     &  15.843 $\pm$  0.029 \\
  7   &  18:57:58.01    &   48:51:49.72     &  16.724 $\pm$  0.027 
\enddata

\tablecomments{Coordinates are J2000 and are based on an astrometric solution obtained by the \textit{astrometry.net} software \citep{lang2010}. The quoted uncertainties are calculated as the standard deviation of 18 measurements from photometric nights at WMO.}
\end{deluxetable}

The steps from performing aperture photometry to obtaining a multiple-telescope light curve were carried out using an automated pipeline. Mapping WCS coordinates onto the images allowed for automatic detection of the AGN and comparison-star locations for aperture photometry. The automated nature of this process enables the pipeline to process a large number of images at once, and to rapidly produce and update the AGN light curve as new images are acquired.

%-----FIGURE-----%

%----------------%

%-----TABLE-----%

\begin{deluxetable}{cccc}
%\tabletypesize{}
%\rotate
\tablewidth{0pt} %0pt is natural width
\tablecolumns{4}
\tablecaption{Photometry measurements for \kagn \label{tab:vbandtable}}
\tablehead{
  \colhead{UT Date}  & 
  \colhead{Telescope}  & 
  \colhead{HJD $-$ 2450000}  & 
  \colhead{\textit{V} (mag)}  
}
\startdata
     2012-01-31    &      N  &    5957.754  &  17.129  $\pm$   0.008  \\
     2012-02-11    &      M  &    5969.014  &  17.012  $\pm$   0.015  \\
     2012-02-17    &      N  &    5974.663  &  17.055  $\pm$   0.015  \\
     2012-02-20    &      F  &    5978.153  &  17.011  $\pm$   0.014  \\
     2012-02-22    &      F  &    5980.166  &  17.037  $\pm$   0.020  

\enddata
\tablecomments{The telescopes are listed as follows: N = Nickel, M = MLO, F = FTN, K = KAIT, W = WMO.
(Full table available in online version.)}
\end{deluxetable}

%---------------------------------------------------------------------------------------------

\subsection{\textit{Kepler} Data}

We also obtained \textit{Kepler} Simple Aperture Photometry (SAP) fluxes for \kagn from the MAST archive for Q13, Q14, and Q15, corresponding to 2012 March through November. Data from Q12 are missing from the archive because, during this time, the source fell on Module 3 of the \textit{Kepler} telescope, which failed early on in the mission.

The \textit{Kepler} light curves are mismatched between individual quarters, so we used our \textit{V}-band light curve as a reference to scale each quarter's light curve individually. We applied a different multiplicative scale factor and additive shift to each quarterly \textit{Kepler} light curve to bring it into agreement with ground-based observations. The multiplicative factors account for the difference in transmission between the \textit{Kepler} and \textit{V}-band filters, and the additive constants account for the changes in AGN-to-host galaxy flux ratio between each quarter caused by using different quarterly extraction apertures to obtain SAP fluxes.

\begin{figure}
%\plotone{f3.eps}
\epsfxsize=8.8cm\epsfbox{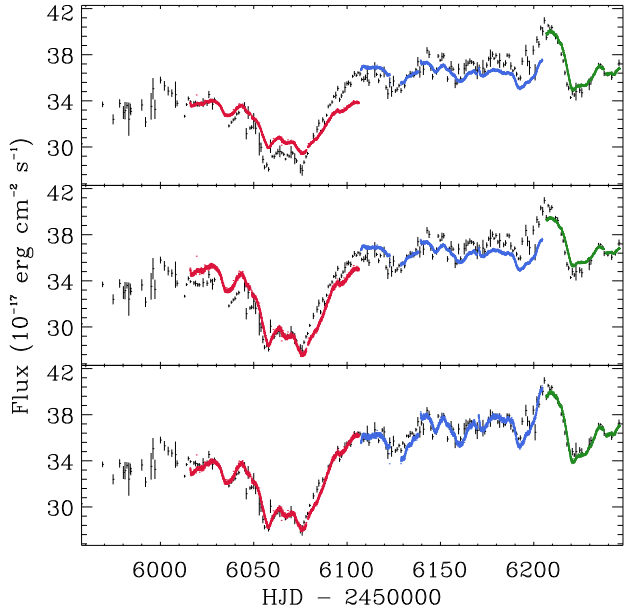}
\caption{\textit{V}-band light curve (black) overplotted with \textit{Kepler} Q13, Q14, and Q15 light curves in red, blue, and green (respectively), scaled by three different methods. \textit{Top:} multiplicative factor only; \textit{middle:} multiplicative factor plus additive constant; \textit{bottom:} multiplicative factor plus additive constant fitted to a \textit{V}-band light curve with an additional linear trend. Error bars are not plotted for the \textit{Kepler} points.}
\label{fig:keplerscaling}
\end{figure}

For each epoch in the condensed \textit{V}-band light curve, we averaged together all \textit{Kepler} flux measurements taken within six hours of the \textit{V}-band measurement to compose condensed \textit{Kepler} light curves. Then for each quarter, we fitted the contemporaneous \textit{Kepler} and $V$-band flux measurements to the equation

\begin{equation}
f_V = m * f_\mathrm{Kepler} + b,
\end{equation}

\noindent where $m$ gives the multiplicative scale factor and $b$ gives the additive shift. We fitted the data using MPFITEXY to account for measurement errors in both \textit{V}-band and \textit{Kepler} data. Figure~\ref{fig:keplerscaling} shows the results of applying a scale factor (top panel) and a scale factor plus a shift (middle panel) to the \textit{Kepler} light curves.

Even with a multiplicative scale factor and an additive shift, however, there are still visible discrepancies between the two sets of data. Specifically, each \textit{Kepler} quarterly light curve tilts downward with time compared to the \textit{V}-band light curve. This is caused by the constant change in \textit{Kepler} pointing with respect to the \textit{Kepler} field as the telescope orbited the Sun, which, in turn, causes differential velocity aberration (DVA) and results in a trend that is superimposed on the light curve within the period of each quarter \citep{stillbarclay2012}. To account for this effect, we applied an empirical secular linear trend to the \textit{V}-band light curve by adding a time-dependent flux to the data. The \textit{Kepler} light curves were then fitted to the adjusted \textit{V}-band light curve with scale factors and shift constants. Finally, the empirical trend was removed from both \textit{V}-band and \textit{Kepler} light curves by subtracting the same time-dependent fluxes as before. The resulting scaled \textit{Kepler} light curves are shown in the bottom panel of Figure~\ref{fig:keplerscaling}, and were used for subsequent \textit{Kepler}-related lag analyses.

%-----FIGURE-----%

%----------------%

We note that the \textit{Kepler} passband is much better matched to the \textit{R} band rather than the \textit{V} band, which means there could be color-dependent variability signals contributing to discrepancies between the \textit{V} and \textit{Kepler} light curves. We also note that the SAP light curves from the \textit{Kepler} archive are susceptible to several instrumental effects. First, the use of different sized apertures between individual quarters affects the SAP fluxes more so than the \textit{Kepler} Pre-search Data Conditioning Simple Aperture Photometry (PDCSAP) fluxes because of the much smaller aperture sizes of SAP. Additionally, the effects of DVA are also larger in the SAP light curves compared to the PDCSAP light curves. A more robust analysis of \textit{Kepler} AGN data would require re-extracting the SAP light curve over a larger set of pixels to remove these systematics, but that is beyond the scope of this work.

%--------------------------------------------------------------------------------

\subsection{Continuum Light Curve Characteristics}

To quantify the observed \kagn continuum variability during our monitoring period, we computed the statistics $R_{\mathrm{max}}$ and $F_{\mathrm{var}}$ for consistency with previous reverberation mapping studies \citep{rodriguez1997,peterson2004}. $R_{\mathrm{max}}$ is defined as the ratio between the maximum and minimum observed fluxes, and $F_\mathrm{var}$ is defined as

\begin{equation}
F_{\mathrm{var}} = \frac{\sqrt{\sigma^2 - \langle\delta^2\rangle}}{\langle f\rangle},
\end{equation}

\noindent where $\sigma^2$ is the sample variance, $\langle \delta^2 \rangle$ is the mean square value of the measurement uncertainties, and $\langle f \rangle$ is the unweighted mean flux. $F_\mathrm{var}$ is essentially an estimate of the intrinsic root-mean-square (rms) variability relative to the mean flux corrected for random errors. We found $R_\mathrm{max} = 1.56$ and $F_\mathrm{var} = 0.086$ for the \textit{V}-band light curve and $R_\mathrm{max} = 1.45$ and $F_\mathrm{var} = 0.080$ for the \textit{Kepler} light curve.

A previous AGN monitoring campaign carried out by the LAMP 2008 collaboration observed 13 AGNs over a two-month period \citep{bentz2009}. Five of these AGNs (Mrk 142, Mrk 1310, Mrk 202, NGC 4253, and NGC 4748) are NLS1 galaxies with full width at half-maximum intensity FWHM$(\hbeta_\mathrm{broad}) < 2000$~\kms. The $F_\mathrm{var}$ values for their \textit{V}-band light curves range from 0.27 to 0.73, and the $R_\mathrm{max}$ values range from 1.12 to 1.39. Compared to these NLS1s, \kagn was significantly more variable during our monitoring period, with both $F_\mathrm{var}$ and $R_\mathrm{max}$ values much larger than those for the LAMP 2008 NLS1 galaxies over their two-month monitoring period.

%====================================================================================================

\section{Spectroscopic Observations}

Spectroscopic observations of \kagn were carried out using the Kast Spectrograph on the Shane 3-m telescope at Lick Observatory. This spectrograph is usually mounted only during dark runs. We employed an interrupt-mode observing method, where every group of Kast observers took one exposure of \kagn on each of their regularly scheduled observing nights. This enabled us to spectroscopically monitor the AGN for a total of nine months, much longer than what is achievable by most dedicated observing campaigns at classically scheduled facilities.

The Kast spectrograph has a D55 dichroic that splits light from the slit at about 5500 \angstrom\ into separate blue- and red-side cameras. Our standard setup used a 600/4310 grism on the blue side, which gives a wavelength dispersion of $1.02$ \angstrom\ \perpix and wavelength range of 2090 \angstrom. However, the wavelength coverage was inconsistent because each group used a slightly different blue-side setup that shifted the wavelength coverage, and on the nights of 2012 February 16, 2012 March 4, 2012 April 19, and 2012 May 1, the observers employed a 830/3460 grism. We used the wavelength range 4000$-$5500 \angstrom\ for our analysis as this is common to all spectra. This wavelength range includes the \hbeta, \hgamma, \hdelta, [\ion{O}{3}], and \ion{He}{2} emission lines, as well as a portion of the Balmer continuum.

\begin{figure}
\epsfxsize=8.75cm\epsfbox{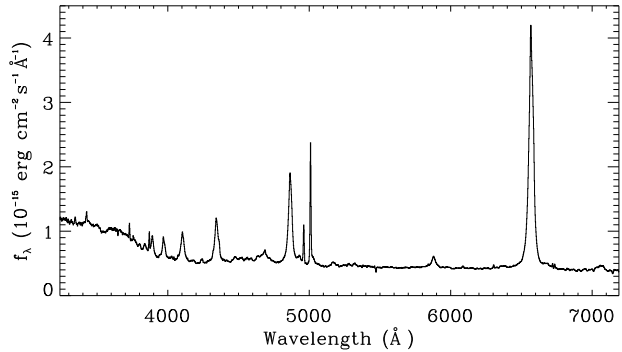}
%\plotone{f4.eps}
\caption{Mean rest-frame spectrum of \kagn constructed from all nights with both blue- and red-side Kast observations.}
\label{fig:fullspectrum}
\end{figure}

On the red side, because different observing teams used significantly different setups for their primary science targets, we were unable to obtain a complete set of spectra with consistent quality and wavelength coverage for analysis of the \halpha line. For reference, Figure~\ref{fig:fullspectrum} shows the unweighted mean AGN spectrum constructed from all nights with both blue- and red-side Kast data.

%-----FIGURE-----%

%----------------%

From February to November of 2012, weather permitting, each regularly scheduled group of Kast observers took at least one 1200 s exposure of \kagn at the parallactic angle \citep{filippenko1982} using a 2\arcsec~slit, along with one 120 s exposure of a flux-standard star with the same slit width for calibration. Two consecutive exposures of \kagn were taken on two nights and three consecutive exposures were taken on four nights. The flux standards we used are BD+284211, Feige 34, G191B2B, and HZ 44, in decreasing order of frequency. Very few spectra were taken in February and March owing to poor weather conditions. April and May had several good nights of data, and starting from June until the end of the campaign in November, we obtained spectra during more than two-thirds of the Kast nights each month. We obtained spectroscopic data from a total of 74 nights.

%====================================================================================================

\section{Spectroscopic Reductions and Measurements}

Spectroscopic data reduction included overscan subtraction, flat fielding, cosmic ray cleaning using the L.A.Cosmic routine \citep{vandok2001}, extraction with a width of $6\farcs88$ (corresponding to a 16-pixel extraction window for the blue-side data), wavelength calibration employing line-lamp exposures, and flux calibration using standard stars. We took unweighted extractions for AGN spectra and optimal extractions for standard-star spectra. Spectra taken on the same night were averaged into a single spectrum. We also propagated the extracted error spectrum through subsequent calibrations and analyses.

\begin{figure}
\epsfxsize=8.70cm\epsfbox{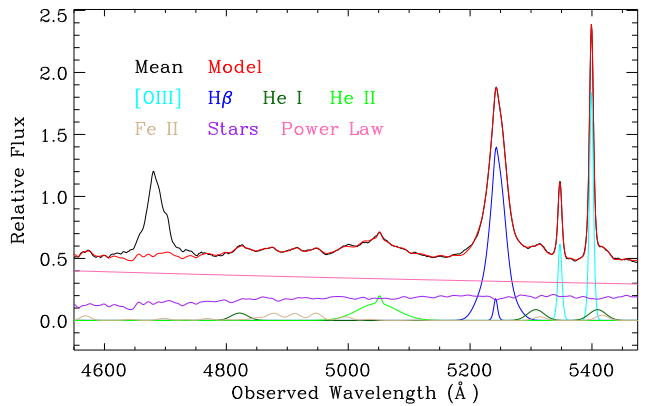}
\caption{\kagn mean spectrum (black), the combined model fit of all components (red), and individual spectral fit components. The \hgamma and [\ion{O}{3}] $\lambda$4363 blend was excluded from the fits in order to limit the total number of fit parameters.}
\label{fig:spectraldecomp}
\end{figure}

We attempted to perform spectral decomposition using methods described by \citet{barth2013} to isolate the broad-line components. However, owing to the presence of weak and blended emission lines as well as limited spectral coverage, many single-epoch spectra produced poorly constrained fit parameters for the continuum components, \ion{He}{2} and \ion{Fe}{2} emission, and reddening. We therefore used the traditional approach of measuring line fluxes by employing a linear fit to approximately subtract the continuum underlying emission lines. The decomposed components of the higher-SNR mean spectrum are displayed in Figure~\ref{fig:spectraldecomp} for reference.

%-----FIGURE-----%

%----------------%

To quantify the flux-measurement uncertainty introduced by using this linear interpolation approach as opposed to the spectral deomposition method, we also measured $f(\hbeta)$ of a series of \hbeta-only spectra. Each \hbeta-only spectrum was created by subtracting from the data all model fit components except the broad and narrow \hbeta models. We found that $f(\hbeta)$ measured from the \hbeta-only spectra are, on average, 1.2\% higher than those measured by simply interpolating over the continuum in the data.

To calibrate the relative fluxes between individual spectra, we followed steps described by \citet{vangro1992}, where the [\ion{O}{3}] lines are taken to be constant in flux for the duration of the campaign. As the [\ion{O}{3}] line is emitted by gas in the narrow-line region, which is much farther out from the black hole than the BLR, the time delay in line response to continuum variations is much longer than typical reverberation mapping campaigns. The algorithm applies a multiplicative flux scaling factor, a small wavelength shift, and a convolution with a Gaussian kernel to a region in each individual spectrum that contains a narrow emission line and some surrounding continuum, and searches for a combination of these parameters that minimizes the residual between this region in the individual spectrum and the same region in a reference spectrum. We constructed the reference spectrum from the mean of all Kast blue-side spectra taken with the 600/4310 grism, and chose the observed wavelength range 5390$-$5410 \angstrom, which encompasses the [\ion{O}{3}] $\lambda5007$ emission line, to be the comparison region. Spectra taken with the 830/3460 grism were not used to make the reference spectrum, but were calibrated using the same method. The flux scale factors range from 0.27 to 4.70. The median wavelength shift is 1.2~\angstrom, which is consistent with the amount expected from miscentering the AGN in the slit.

We followed steps described by \citet{barth2011} to assess the accuracy of the spectral scaling, and calculated the normalized excess variance of the [\ion{O}{3}] emission-line light curve. The normalized excess variance, $\sigma^{2}_{\mathrm{nx}}$, is defined by normalizing Eq.~\ref{eq:excessvariance} by a factor of the mean flux squared, giving

\begin{equation}
\sigma^{2}_{\mathrm{nx}} = \frac{1}{N\mu^{2}} \displaystyle\sum\limits_{i=1}^{N} [(X_{i} - \mu)^{2} - \sigma^{2}_{i}].
\end{equation}

\noindent We found $\sigma_{\mathrm{nx}} \approx 0.02$ for the [\ion{O}{3}] light curve after flux scaling, indicating that, above the uncertainties from photon counting in flux measurements, there is an additional scatter on the order of $2\%$ of the total [\ion{O}{3}] flux in the scaled light curve. This scatter may be caused by a combination of variations in seeing, miscentering of the AGN in the slit, and nightly variations in the instrument focus. Overall, this is a relatively small effect on the flux scaling of the \hbeta light curve. We added this $2\%$ flux scatter in quadrature to all spectroscopic flux uncertainties before performing further analysis.

The spectroscopic data were photometrically calibrated by carrying out synthetic \textit{V}-band photometry on the spectrum from 2012 September 9, which was taken under nearly photometric conditions. We compared this magnitude to the aperture photometry magnitude from the same night and calculated a scale factor of 1.15 that needed to be applied to the spectrum to bring the synthetic photometry measurement into agreement with the aperture photometry measurement. We then applied this scale factor to the entire set of Kast spectra.

\begin{deluxetable}{lcc}
%\tablewidth{0pt} 
\tablewidth{8.6cm} 
\tablecolumns{4}
\tablecaption{Wavelength Windows for Flux Measurements \label{tab:wavelenwindows}}
\tablehead{
\colhead{Line (\angstrom)} & \colhead{Line Window (\angstrom)} & \colhead{Continuum Windows (\angstrom)}
}
\startdata
\hbeta       & 5200$-$5290   & 5130$-$5160, 5360$-$5390    \\
\ion{He}{2}  & 4990$-$5100   & 4960$-$4980, 5120$-$5160    \\
\hgamma      & 4650$-$4720   & 4600$-$4640, 4730$-$4780    \\
\hdelta      & 4395$-$4455   & 4360$-$4380, 4470$-$4500    \\
\hbeta-blue  & 5200$-$5238   & 5130$-$5160, 5360$-$5390    \\
\hbeta-core  & 5239$-$5249   & 5130$-$5160, 5360$-$5390    \\
\hbeta-red   & 5250$-$5290   & 5130$-$5160, 5360$-$5390    
\enddata

\tablecomments{Wavelengths are in the observed frame.}
\end{deluxetable}

\begin{deluxetable*}{ccccccc}
%  \tablewidth{0pt} 
  \tablecaption{Spectroscopic Measurements for \kagn \label{tab:spectable}}
  \tablecolumns{6}
  \tablehead{
    \colhead{UT Date} & 
    \colhead{HJD$-$2450000} & 
    \colhead{SNR} & \colhead{$f$(\hbeta)} & 
    \colhead{$f$(\hgamma)} & 
    \colhead{$f$(\hdelta)} & 
    \colhead{$f$(\heii)} \\
    \colhead{} &
    \colhead{} &
    \colhead{} &
    \multicolumn{4}{c}{(\fluxunit)}
  }
\startdata
    2012-02-16   &    5974.087   &   11   & 42.87   $\pm$ 0.38   &  19.12  $\pm$ 0.51   &  \phn8.27 $\pm$ 0.54   &     11.60 $\pm$ 0.47    \\
    2012-03-04   &    5991.073   &   35   & 40.90   $\pm$ 0.14   &  18.37  $\pm$ 0.14   &  \phn9.84 $\pm$ 0.14   &  \phn7.30 $\pm$ 0.14    \\
    2012-04-02   &    6020.024   &   19   & 42.42   $\pm$ 0.30   &  19.30  $\pm$ 0.27   &     12.24 $\pm$ 0.27   &  \phn9.68 $\pm$ 0.28    \\
    2012-04-16   &    6033.929   &   21   & 45.20   $\pm$ 0.28   &  22.75  $\pm$ 0.26   &     13.73 $\pm$ 0.26   &     10.20 $\pm$ 0.27    \\
    2012-04-16   &    6034.975   &   34   & 46.41   $\pm$ 0.19   &  22.53  $\pm$ 0.16   &     13.15 $\pm$ 0.16   &  \phn8.04 $\pm$ 0.16    

\enddata
\tablecomments{Listed SNR is the signal-to-noise ratio per pixel for the observed wavelength range 4500--4600 \angstrom\ in the AGN spectra. Measured fluxes include the blended broad and narrow emission lines.
(Full table available in online version.)}
\end{deluxetable*}

To obtain emission-line fluxes, we first subtracted a local linear continuum surrounding the line, then integrated over the emission-line profile. Table~\ref{tab:wavelenwindows} shows the wavelength ranges used for each line and their local continuum windows. Table~\ref{tab:spectable} gives the spectroscopic measurements of the \hbeta, \hgamma, \hdelta, and \ion{He}{2} emission lines for the entire dataset, as well as the SNR for each epoch measured using the observed wavelength range 4500$-$4600 \angstrom. The median SNR per pixel is 28.

\begin{figure*}[h]
\centering
%\plotone{f6.eps}
%\scalebox{0.5}{\includegraphics{f6.eps}}
\epsfxsize=16cm\epsfbox{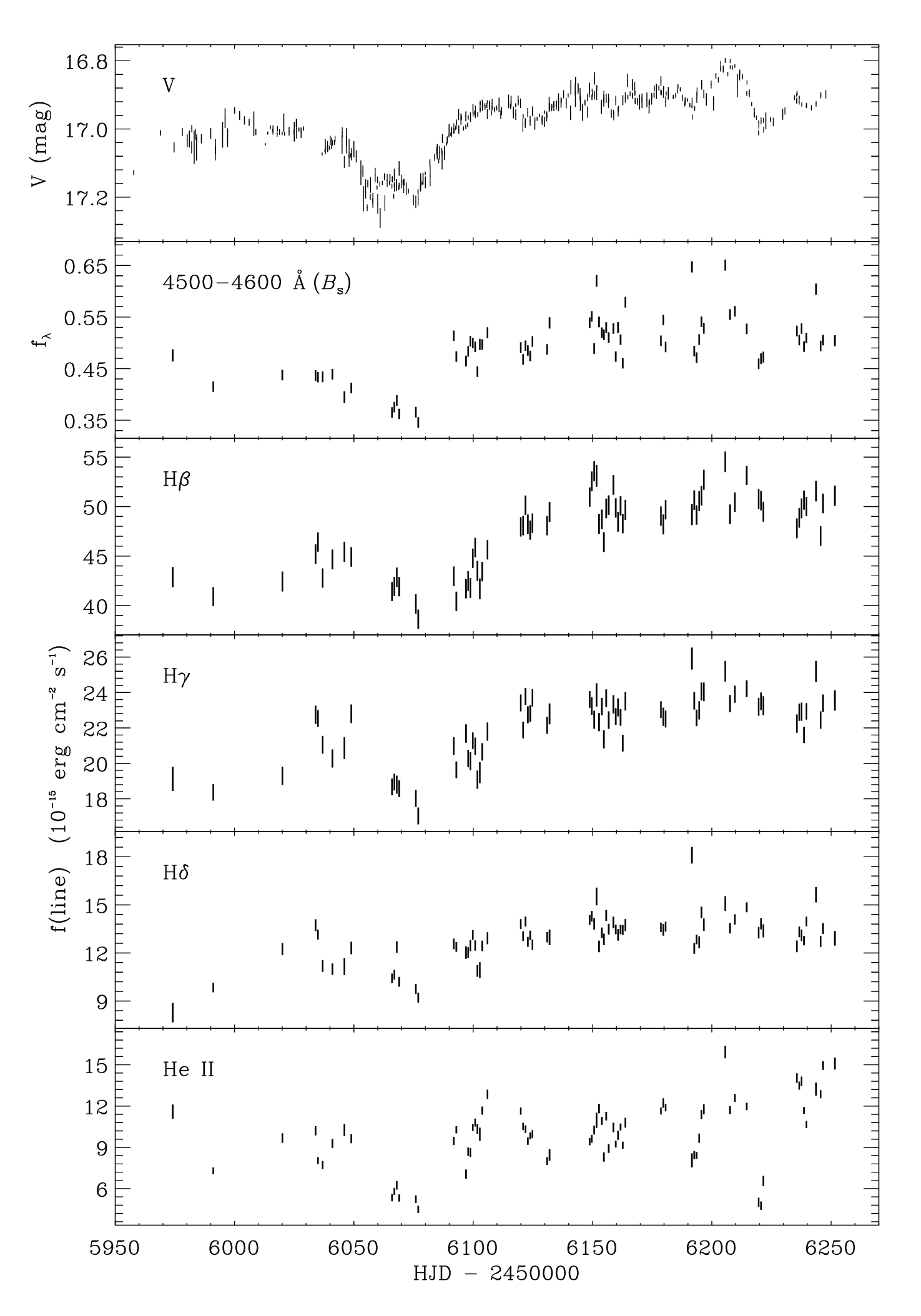}
\caption{\kagn \textit{V}-band magnitude, continuum flux measured from the spectroscopic data, and emission-line light curves. Plotted errors include the 2\% flux scatter found by computing the normalized excess variance of the [O III] light curve.}
\label{fig:multiltcurve}
\end{figure*}

Figure~\ref{fig:multiltcurve} displays the \textit{V}-band photometric and spectroscopic light curves for the \hbeta, \hgamma, \hdelta, and \ion{He}{2} emission lines. The scaling routine works best for wavelength ranges closest to the [\ion{O}{3}] emission lines, so at wavelengths farther away from [\ion{O}{3}], the higher-order Balmer-line light curves become progressively noisier. Noise in the \ion{He}{2} light curve is primarily caused by weak line strength as well as a lack of true continuum surrounding the line. The presence of \ion{Fe}{2} lines blended into the blue side of \ion{He}{2}, and the fact that the \ion{He}{2} line is intrinsically very weak and broad, make fitting the true continuum with a linear model very difficult. The spectral decomposition components of \ion{He}{2} are also poorly constrained owing to the line's low amplitude.

Figure~\ref{fig:multiltcurve} also illustrates the spectroscopic light curve for the observed wavelength range 4500$-$4600 \angstrom. This region is dominated by continuum emission, so its light curve can be compared with the \textit{V}-band light curve. This spectroscopic \textit{B}-band continuum light curve, denoted by $B_\mathrm{s}$, is noisier than that of the \textit{V} band owing to higher susceptibility to seeing variations and slit losses, but the two light curves show consistent variability trends during the monitoring period.

The $F_{\mathrm{var}}$ and $R_{\mathrm{max}}$ values for each of the light curves are listed in Table~\ref{tab:ltcurve_stats}. The higher-order Balmer lines exhibit distinctly larger relative variability amplitude, and the \ion{He}{2} line is proportionally more variable than all the Balmer lines. Both results are in agreement with findings of previous reverberation mapping programs \citep{petersonferland1986,dietrich1993,kollatschny2003,bentz2010}.

\begin{deluxetable}{lcc}[h]
%\tablewidth{0pt}
\tablewidth{8.5cm} 
\tablecolumns{3}
\tablecaption{Light Curve Statistics \label{tab:ltcurve_stats}}
\tablehead{
\colhead{Light Curve} & \colhead{$F_{\mathrm{var}}$} & \colhead{$R_{\mathrm{max}}$}
}
\startdata
\textit{V}       &  0.084  & 1.50 $\pm$ 0.05   \\
$B_\mathrm{s}$     &  0.119  & 1.88 $\pm$ 0.07  \\
\hbeta           &  0.076  & 1.41 $\pm$ 0.05  \\
\hgamma          &  0.078  & 1.52 $\pm$ 0.07   \\
\hdelta          &  0.111  & 2.19 $\pm$ 0.13   \\
\ion{He}{2}      &  0.245  & 3.54 $\pm$ 0.26    

\enddata

\tablecomments{$R_{\mathrm{max}}$ and $F_{\mathrm{var}}$ values for \textit{V}, $B_\mathrm{s}$, and the four emission lines. Higher-ionization lines show larger variations.}

\end{deluxetable}

\begin{figure}
%\plotone{f7.eps}
\epsfxsize=8.75cm\epsfbox{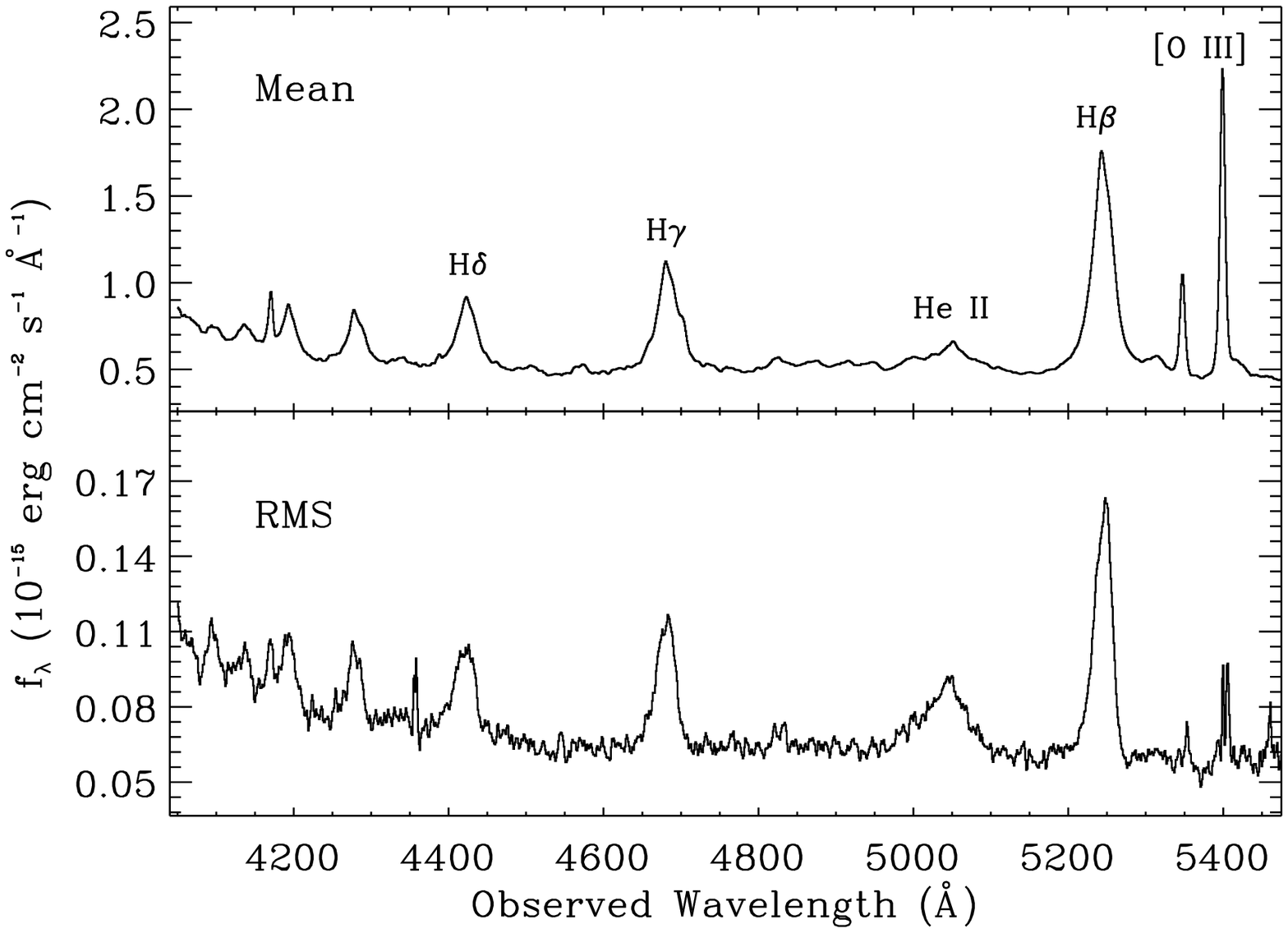}
\caption{Mean and rms spectra of \kagn.}
\label{fig:meanrmsspectrum}
\end{figure}

Figure~\ref{fig:meanrmsspectrum} shows the mean and rms spectra of \kagn constructed from all blue-side spectra taken with the 600/4310 grism after applying [\ion{O}{3}] spectral scaling. The rms spectrum indicates the amount of relative variability at each wavelength. The [\ion{O}{3}] narrow lines have low residuals in the rms spectrum, indicating good spectral flux calibration results using the [\ion{O}{3}] lines. The broad Balmer lines clearly stand out with very high variability. \ion{He}{2} appears to be highly variable in the rms spectrum, even though the line is weak in the mean spectrum owing to blending with \ion{Fe}{2} lines.

%-----FIGURE-----%

%---------------%

%-----FIGURE-----%

%---------------%

%====================================================================================================

\section{Lag Measurements}
\label{sec:lag}

%------------------------------------------------------------------------------------------

\subsection{Cross-Correlation Measurements}

We calculated the lag between the continuum and each emission-line light curve illustrated in Figure~\ref{fig:multiltcurve}, as well as between the photometric and spectroscopic light curves, by employing the interpolation cross-correlation technique developed by \citet{gaskellpeterson1987} and described by \citet{whitepeterson1994}, \citet{peterson2004}, and \citet{bentz2009}. We computed the cross-correlation function (CCF) for $\tau$ values from $-$20 to 40 days in increments of 0.25 days. The lag for each emission line is then calculated in two ways: by using the peak of the CCF, defined as \taupeak, and by using the centroid of CCF values above 80\% of the peak value, defined as \taucen. We opted to use \taucen for \mbh estimates as \citet{peterson2004} showed that this yields more consistent black hole mass estimates between different emission lines.

\begin{figure}
%\plotone{f8.eps}
\epsfxsize=8.65cm\epsfbox{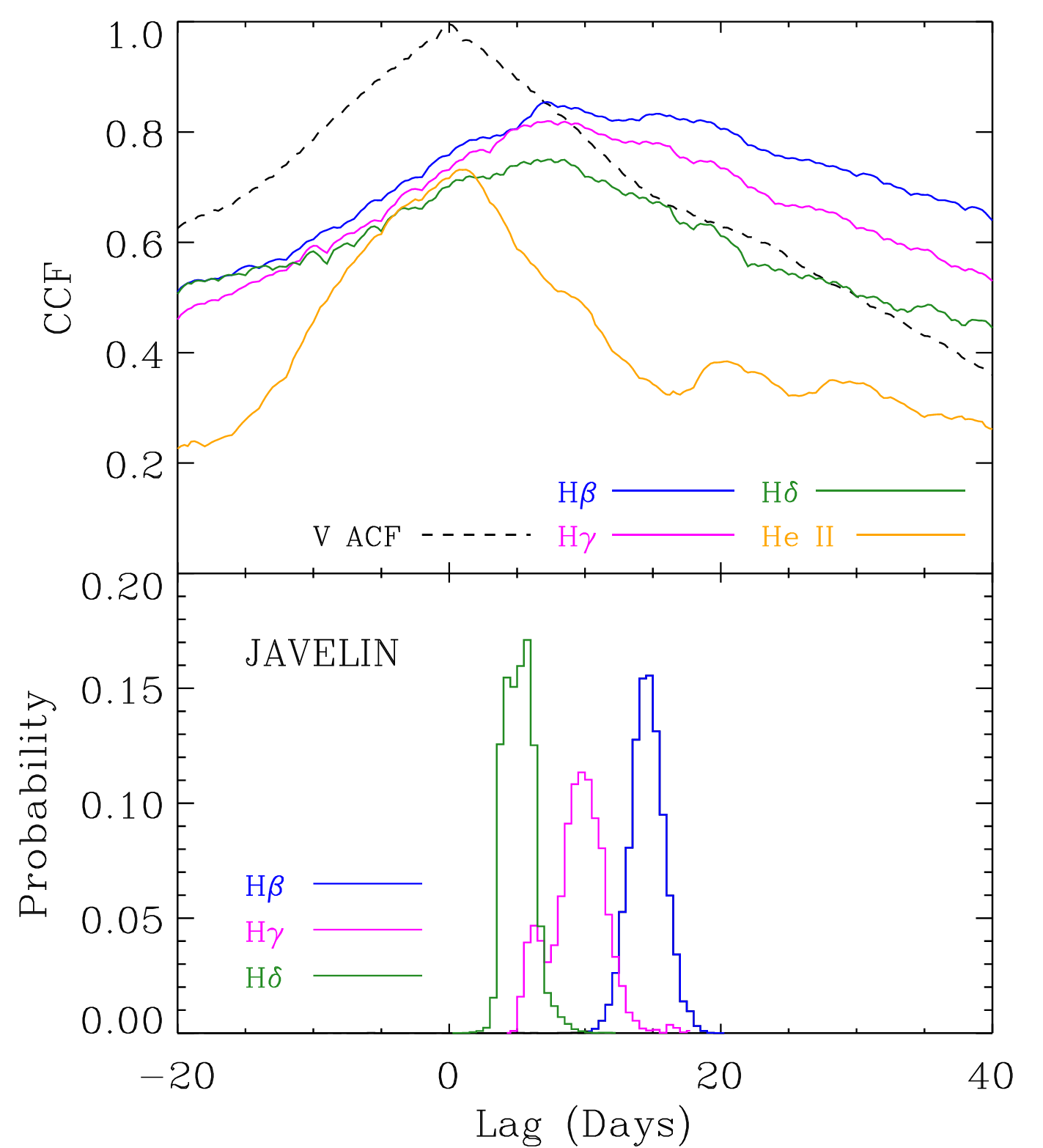}
\caption{\textit{Top}: Cross-correlation functions between the four emission lines and the \textit{V}-band continuum, and the auto-correlation function of the \textit{V}-band continuum. \textit{Bottom}: Probability distributions of \texttt{JAVELIN} lags for \hbeta, \hgamma, and \hdelta. These distributions were obtained with $6.25 \times 10^4$ iterations, while the \ion{He}{2} distribution was obtained with $2.5 \times 10^5$ iterations. However, the \ion{He}{2} distribution is poorly constrained owing to the line's noisy light curve, and is therefore omitted in this plot.}
\label{fig:multiccf_javelin}
\end{figure}

In cases where the continuum light curve exhibits distinct global trends, a detrending procedure is sometimes applied prior to cross-correlation analysis, where a linear function is fitted to and subtracted from the light curve so that only local variations are taken into account in the cross-correlation. We computed the \hbeta lag both with and without detrending using a linear fit. In the case without detrending, the lag uncertainties are smaller and the CCF peak is higher, indicating a more robust CCF. Therefore, we chose to omit the detrending procedure for our final cross-correlation analysis. The top panel in Figure~\ref{fig:multiccf_javelin} shows the CCF for the four emission-line light curves with the photometric light curve. We also computed the auto-correlation function (ACF) for the photometric light curve, which peaks at zero lag as expected.

%-----FIGURE-----%

%---------------%

To determine the final lags and their uncertainties, we employed the same Monte Carlo bootstrapping method used by \citet{barth2011} and described by \citet{whitepeterson1994} and \citet{peterson2004}. We constructed $10^4$ modified realizations of the continuum and emission-line light curves. Each realization is made by randomly choosing $n$ data points from the actual light curve allowing resampling, where $n$ is the total number of points in the dataset. If a point is picked $m$ times, then its uncertainty is reduced by a factor of $m^{1/2}$. The simulated light curves are then varied by adding random Gaussian noise based on the measured uncertainties at each data point. We then computed the CCF for each pair of simulated continuum and line light curves to construct distributions of \taucen values. The median values are chosen as the final lag results, and the uncertainties on \taucen are the 1$\sigma$ thresholds in the distribution centered around the median.

Table~\ref{tab:lagtable} gives the measured \taupeak and \taucen values for the four emission-line light curves with respect to the \textit{V}-band light curve. The \ion{He}{2} lag is consistent with zero within 1$\sigma$ uncertainties. The larger fractional uncertainties on the higher-order Balmer line lags, as well as on the \ion{He}{2} lag, can be attributed to their noisier light curves due to less precise spectral scaling at wavelengths farther from [\ion{O}{3}].

The lag times are progressively shorter for higher-order Balmer lines. Specifically, we find lag ratios of $\tau$(\hbeta):$\tau$(\hgamma):$\tau$(\hdelta)$=$1.00:0.75:0.44. This is consistent with the picture of a BLR stratified in optical depth \citep{rees1989,koristagoud2004}, as well as with findings from previous reverberation mapping campaigns (e.g., \citealt{bentz2010}).

%-----TABLE-----%

\begin{deluxetable}{lccc}
\tablecaption{Observed-Frame Lag Measurements \label{tab:lagtable}}
\tablecolumns{4}
%\tablewidth{0pt}
\tablewidth{8.6cm}
\tablehead{
\colhead{Emission Line} & 
\colhead{\taupeak~(days)} & 
\colhead{\taucen~(days)} & 
\colhead{\taujav (days)}
}
\startdata

\phs\hbeta vs. $V$               &  \phs\phn$8.25^{+7.25}_{-1.00}$     &  \phs{$14.58^{+2.19}_{-2.50}$}    &  \phs{$14.18^{+1.16}_{-1.08}$}   \\
\phs\hgamma vs. $V$              &  \phs\phn$7.50^{+2.00}_{-1.25}$     &  \phs{$10.96^{+3.08}_{-2.76}$}    &  \phs{$10.04^{+1.42}_{-1.48}$}   \\
\phs\hdelta vs. $V$              &  \phs\phn$6.50^{+1.75}_{-2.00}$     &  \phn\phs{$6.42^{+2.53}_{-2.60}$} &  \phn\phs{$5.81^{+1.06}_{-2.03}$}   \\
\phs \ion{He}{2} vs. $V$         &  \phs\phn$0.75^{+0.50}_{-0.50}$     &  \phn$-0.58^{+1.20}_{-0.85}$      &  \phn$-2.86^{+2.01}_{-0.08}$   \\
\phs\hbeta blue vs. $V$          &  \phs\phn$7.25^{+1.00}_{-0.75}$     &  \phs$13.43^{+2.17}_{-2.62}$      &  \phs$13.85^{+1.22}_{-1.23}$    \\
\phs\hbeta core vs. $V$          &  \phs$15.00^{+3.75}_{-6.50}$        &  \phs$15.50^{+1.92}_{-2.01}$      &  \phs$14.73^{+0.90}_{-0.89}$   \\
\phs\hbeta red vs. $V$           &  \phs\phn$8.50^{+5.75}_{-1.25}$     &  \phs$12.89^{+3.64}_{-3.20}$      &  \phs$14.25^{+1.28}_{-1.26}$    \\
\phs\hbeta vs. $B_\text{s}$       &  \phs$11.50^{+5.50}_{-4.00}$        &  \phs$14.89^{+4.19}_{-5.10}$      &  \phs$15.67^{+1.20}_{-1.62}$     \\
\phs $V$ vs. $B_\text{s}$         &  \phs\phn$2.25^{+1.25}_{-2.75}$     &  \phs\phn$1.68^{+2.21}_{-1.39}$   &  \phs\phn$1.64^{+0.30}_{-0.73}$  \\
\phs\hbeta vs. \textit{Kepler}          &  \phs\phn$8.25^{+6.50}_{-1.00}$     &  \phs$14.17^{+2.26}_{-2.66}$      &  \phs$13.42^{+1.10}_{-1.10}$  \\
\phs\hgamma vs. \textit{Kepler}         &  \phs\phn$6.75^{+1.50}_{-1.25}$     &  \phs\phn$9.49^{+3.02}_{-2.24}$   &  \phn\phs$9.10^{+0.93}_{-0.89}$  \\
\phs\hdelta vs. \textit{Kepler}         &  \phs\phn$4.50^{+1.75}_{-1.75}$     &  \phn\phs$4.86^{+2.78}_{-2.27}$   &  \phn\phs$4.86^{+0.86}_{-0.73}$      \\
\phs \ion{He}{2} vs. \textit{Kepler}    &  \phn\phs$0.00^{+0.50}_{-0.75}$     &  \phn$-0.72^{+0.72}_{-0.72}$      &  \phn\phs$0.88^{+0.03}_{-0.03}$   \\
\phs \textit{Kepler} vs. $V$            &  \phs\phn$0.50^{+0.25}_{-0.00}$     &  \phs\phn$1.00^{+0.47}_{-0.47}$   &  \phn\phs$0.76^{+0.31}_{-0.30}$  \\
\phs \textit{Kepler} vs. $B_\text{s}$    &  \phs\phn$1.75^{+1.50}_{-1.25}$     &  \phs\phn$1.95^{+1.28}_{-1.16}$   &  \phn\phs$2.06^{+0.15}_{-2.15}$

\enddata
\tablecomments{Cross-correlation $\tau_\mathrm{peak}$, cross-correlation $\tau_\mathrm{cen}$, and \texttt{JAVELIN} lags. Observed-frame lags can be converted to rest-frame lags by dividing by $1+z$.}
\end{deluxetable}

%---------------%

Additionally, we attempted to obtain velocity-resolved lag measurements for \kagn, since the lag behavior as a function of velocity across broad emission lines can contain information about BLR kinematics. We divided the \hbeta line profile into three wavelength segments: 5200$-$5238~\angstrom\ for the blue wing, 5239$-$5249~\angstrom\ for the core, and 5250$-$5290~\angstrom\ for the red wing. The \hbeta lag for each segment is listed in Table~\ref{tab:lagtable}. We found marginal evidence for longer lag in the emission-line core and shorter lags in the wings. We were unable to obtain useful lag measurements for smaller velocity bins, and therefore refrain from drawing any definitive conclusions regarding the kinematics of the BLR.

%------------------------------------------------------------------------------------------

\subsection{\texttt{JAVELIN}}

We used an alternative method of estimating emission-line lags, which employs a statistical model for quasar variability. This method uses the \textit{Python} code \texttt{JAVELIN} v.0.3$\alpha$ \citep{zu2011} to model the optical AGN continuum variability as a damped random walk process with covariance function 

\begin{equation}
S_{\mathrm{DRW}}(\Delta t) = \sigma^2 \exp\left(-\left|\frac{\Delta t}{\tau_\mathrm{r}}\right|\right),
\end{equation}

\noindent where $\tau_\mathrm{r}$ is the ``relaxation time'' required for the variability to become roughly uncorrelated, and $\sigma$ is the variability amplitude on timescales much shorter than $\tau_\mathrm{r}$ \citep{kelly2009}. \texttt{JAVELIN} fits $\tau_\mathrm{r}$ and $\sigma$ for the AGN continuum light curve, then models the emission-line light curves as lagged, smoothed, and scaled versions of the continuum light curve. An important caveat of using \texttt{JAVELIN} for the \kagn lag analysis is that the \citet{kelly2009} damped random walk model produces variability power spectra with a slope of $-2$, while \citet{mushotzky2011} showed that \kagn has a power-spectrum slope of $\sim -3$.

The \textit{V}-band light curve was rebinned into one-day intervals for analysis with \texttt{JAVELIN} in order to cut down on computation time. While \texttt{JAVELIN} is, in principle, able to fit a large number of emission-line light curves simultaneously, the lags were poorly constrained in this case for fitting three emission-line light curves simultaneously, most likely because of the monthly gaps in the data when the Moon was bright. Therefore, we chose the two-line analysis method, where we fit each of \hgamma, \hdelta, and \heii emission-line light curves simultaneously with that of \hbeta. The \hbeta lags computed from pairing with \hgamma and \hdelta are consistent with each other, while the \hbeta lag computed from pairing with \heii yielded a slightly shorter lag. This is likely due to the noisy \heii light curve as well as the fact that \heii intrinsically has a lag that is very short compared to the monthly gaps in the light curves, which makes the lag difficult to measure. We use the \hbeta lag value obtained from pairing with \hgamma as $\tau_\mathrm{\texttt{JAVELIN}}$ for \hbeta.

\begin{figure}
\epsfxsize=8.95cm\epsfbox{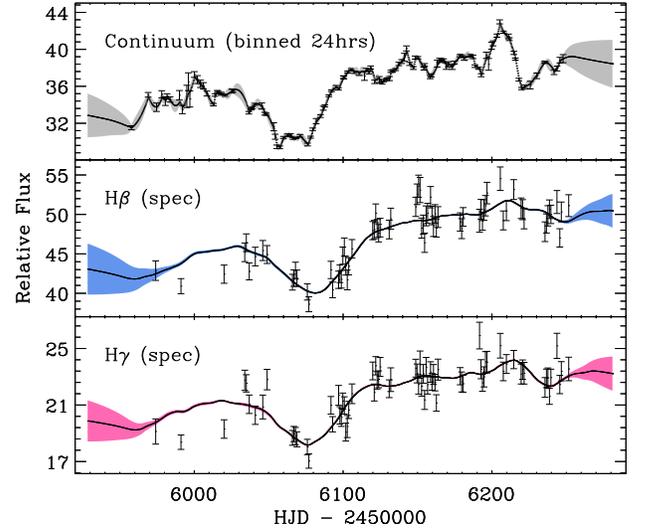}
%\plotone{f9.eps}
\caption{\texttt{JAVELIN} model results for the continuum (\textit{V} band), \hbeta (spectroscopic), and \hgamma (spectroscopic) light curves (black solid lines), the model $1\sigma$ uncertainties at each time (shaded regions), and observational data and uncertainties.}
\label{fig:javelin}
\end{figure}

Table~\ref{tab:lagtable} lists the \texttt{JAVELIN} lags, which are consistent with those obtained using cross-correlation techniques within 1$\sigma$ uncertainties. Lags for the \hbeta blue wing, core, and red wing were computed simultaneously in a three-line \texttt{JAVELIN} run, and the \textit{V}-band and \hbeta lags with respect to the $B_\mathrm{s}$ band were obtained from a two-line run. The bottom panel of Figure~\ref{fig:multiccf_javelin} shows the \texttt{JAVELIN} distributions for \hbeta, \hgamma, and \hdelta lags, and Figure~\ref{fig:javelin} shows the \texttt{JAVELIN} model results for the continuum, \hbeta, and \hgamma light curves.

We note that both the CCF and \texttt{JAVELIN} \ion{He}{2} lags are slightly negative, which is likely caused by the combined effects of the higher ionization (and therefore shorter lag) of \ion{He}{2}, and a slight contaminating lag signal in the \textit{V}-band light curve, described in the next section.

%----------FIGURE-------------

%------------------------------------------------------------------------------------------

\subsection{\hbeta Contamination in the \textit{V} band}

There is a small biasing factor in the \hbeta lag calculations from the emission-line contribution to the \textit{V}-band flux. While the \textit{V}-band light curve would ideally represent pure continuum, the presence of the \hbeta line in the \textit{V} filter adds a flux component that contains a lag signal. Consequently, the calculated lags from the biased continuum would be shorter than those obtained with a pure continuum.

To determine the magnitude of this contribution, we first combined the blue- and red-side spectra from a single night, and next performed synthetic photometry on the spectrum using a Johnson \textit{V} filter. We then removed the \hbeta line from the spectrum by directly interpolating over it, performed synthetic photometry on the modified spectrum, and compared the two magnitude results. We found that \hbeta contributes approximately 9.6\% of the \textit{V}-band flux, and assume that \hbeta dominates the variable emission-line contribution in the \textit{V} band and is therefore the main source of the lag bias.

To quantify the effect of this bias on the calculated lag, we simulated $10^4$ pure AGN continuum light curves using methods described by \citet{timmerkonig1995}, and simulated corresponding emission-line light curves by convolving the continuum light curves with a $\delta$ function at a lag of 14 days. We then simulated $10^4$ contaminated \textit{V}-band light curves by adding a lagged emission-line contribution to the continuum at the 9.6\% level. The pure and contaminated continuum light curves in each pair are both then cross-correlated with the corresponding emission-line light curve to create two distributions of lag times. We found a median lag of $14.0 \pm 2.1$ days for the pure continuum case and a median lag of $13.2 \pm 2.1$ days for the contaminated continuum case, indicating an expected bias of 0.8 days. This prediction is similar to the bias we find from observations.

We found an \hbeta lag of $14.89^{+4.19}_{-5.10}$ days with respect to $B_\mathrm{s}$, which contains no \hbeta flux contamination, indicating an observed bias of $\sim$0.3 days compared to the lag-contaminated case. We also found a small positive lag of $1.68^{+2.21}_{-1.39}$ days for the \textit{V} light curve with respect to $\textit{B}_\mathrm{s}$. However, in both simulations and observations, the biases are smaller than the lag uncertainties for \hbeta with respect to the \textit{V}-band light curve. We therefore conclude that the lag bias due to \hbeta flux contribution in the \textit{V} band is present, but is small compared to the $1\sigma$ uncertainties on the \taucen measurements.

%------------------------------------------------------------------------------------------

\subsection{Lags with Respect to \textit{Kepler} Light Curve}

We computed the lag of each emission line with respect to the scaled \textit{Kepler} fluxes using both cross-correlation analysis and \texttt{JAVELIN}. The \textit{Kepler} light curve has a cadence of 30 minutes, giving a total of $\sim 1.3 \times 10^4$ data points over three quarters. We binned the light curve into bins of 12 and 24 hours to use in the cross-correlation and \texttt{JAVELIN} analyses, respectively. For both CCF and \texttt{JAVELIN}, we found the emission-line lags with respect to the \textit{Kepler} light curve, listed in Table~\ref{tab:lagtable}, to be consistent with but slightly shorter than those with respect to the \textit{V}-band continuum. This is consistent with expectations, since the \textit{Kepler} passband includes \halpha, which introduces an additional lag signal to the \textit{Kepler} light curve compared to the \textit{V}-band data. The redder portion of the continuum could also have a small lag with respect to the bluer continuum \citep{sergeev2005}, since the redder continuum emission comes from larger radii in the accretion disk than where the \textit{V}-band continuum is emitted. The combined effects of broad emission lines and red continuum in the \textit{Kepler} band should account for the shorter emission-line lags measured against the \textit{Kepler} light curve as compared to those measured against the \textit{V}-band light curve.

We also measured the lag of the \textit{Kepler} light curve with respect to both the \textit{V}-band and $B_\text{s}$-band light curves, and found small positive lags for both cases. This also supports the idea of broad emission lines and the red continuum introducing a lag signal to the \textit{Kepler} light curve.

%============================================================================================================

\section{Line Widths and Black Hole Mass Estimate}

There are two conventional methods of measuring the broad-line width: using the FWHM and the line dispersion ($\sigma_\mathrm{line}$) of the emission-line profile. The line profile is typically taken to be the rms profile, since using the variable portion of the spectrum instead of the mean spectrum implies a black hole mass estimate based only on components of the emission line that echo the continuum signal \citep{peterson2004}. The line dispersion is defined as 

\begin{equation}
\sigma^{2}_{\mathrm{line}} = \Bigg(\frac{c}{\lambda_0}\Bigg)^2 \Bigg(\frac{\sum \lambda^2_iS_i}{\sum S_i} - \lambda^2_0 \Bigg),
\end{equation}

\noindent where $S_i$ is the flux density at wavelength bin $\lambda_i$ and $\lambda_0$ is the flux-weighted centroid wavelength of the line profile. In this empirical method of measuring the line width, the line profile is not fitted to any functional model. We used the same line and continuum windows to measure the line width as those used in measuring line fluxes.

To determine the final FWHM and $\sigma_{\mathrm{line}}$ values and uncertainties, we employed the bootstrap method described by \citet{peterson2004}. The entire dataset contains $N$ spectra. For each bootstrap realization, we randomly selected $N$ spectra from the dataset allowing reselection, constructed the mean and rms line profiles from this randomly sampled set, and measured the line dispersion of the rms profile. From multiple realizations, we built up a distribution of FWHM and line-dispersion values, and took the median and standard deviation of the distributions to be the final FWHM and $\sigma_\mathrm{line}$ and their uncertainties, respectively. We removed the instrumental line width by taking the width of the $\lambda$5086 \ion{Cd}{1} calibration line in a 2\arcsec-slit width exposure and subtracting it from the measured FWHM or $\sigma_\mathrm{line}$ in quadrature. We found $\mathrm{FWHM} = 324$ \kms for the \ion{Cd}{1} calibration line for a Gaussian fit to the line profile. After correcting for the instrumental line width, we found $\mathrm{FWHM} = 1511 \pm 68$ \kms and $\sigma_{\mathrm{line}} = $ \hbetawidth for the \hbeta line in the rms spectrum.

We also measured the \hbeta FWHM and $\sigma_\mathrm{line}$ for the mean profile. To ensure exclusion of the narrow-line component in the width measurements, we measured the FWHM and $\sigma_\mathrm{line}$ of the broad \hbeta model based on the spectral decomposition of the mean spectrum, as shown in Figure~\ref{fig:spectraldecomp}. The [\ion{O}{3}] narrow-line profile was used to model the narrow \hbeta line in the spectral fitting routines, and $f(\hbeta_\mathrm{narrow})/f([$\ion{O}{3}$]_\mathrm{\lambda5007}$) $\approx$ 0.09. We also measured the FWHM and $\sigma_\mathrm{line}$ of the broad \hbeta model for each epoch in our dataset, and took the standard deviations about the means to be the FWHM and $\sigma_\mathrm{line}$ uncertainties. We found $\mathrm{FWHM} = 1820 \pm 79$ \kms and $\sigma_{\mathrm{line}} = 853 \pm 34$ \kms for \hbeta in the mean spectrum after subtracting the instrumental line width. This is consistent with previous findings that line widths measured from mean spectra tend to be larger than those measured from rms spectra (e.g., \citealt{bentz2009}).

The reverberation lag and line width of \hbeta combined can give a virial estimate of the central black hole mass, given by

\begin{equation}
M_{\mathrm{BH}} = f \frac{(c\tau)(\Delta V)^2}{G},
\end{equation}

\noindent where $\tau$ is the \hbeta lag time with respect to the continuum and $c\tau$ gives the mean radius of the BLR, $\Delta V$ is the \hbeta line width, $G$ is the gravitational constant, and $f$ is a scaling factor of order unity that depends on the inclination and kinematics of the BLR. Traditionally, since these properties of the BLR are usually unknown, the scale factor $f$ is chosen to be a value that brings the set of reverberation mapped AGNs into agreement with local quiescent galaxies in the \msigma relation, which relates black hole mass to host-galaxy bulge stellar velocity dispersion \citep{onken2004,woo2010,park2012,grier2013}.

We use $c$\taucen for the BLR radius and \hbeta line dispersion $\sigma_{\mathrm{line}}$ of the rms line profile for $\Delta V$, for consistency with \citet{peterson2004}, and a scale factor of $f = 5.13$, calculated by \citet{park2012} based on the updated local AGN \msigma relation obtained with the forward regression method. Combining the \hbeta vs. \textit{V} lag of \lag days, corresponding to a rest-frame lag of $\taucen = \lagrest$ days, and $\sigma_{\mathrm{line}} =$ \hbetawidth, we obtain a virial black hole mass estimate of \bhmass. If we follow the prescription of \citet{grier2013} by using the \hbeta $\tau_\mathrm{\texttt{JAVELIN}}$ and a scale factor $f = 4.31$, we find \bhmassjavelin.

The above uncertainties on $M_{\mathrm{BH}}$ include only errors propagated from the lag and emission-line-width measurements. If we incorporate the uncertainties on the mean scale factor from the linear fits by \citet{park2012}, $f = 5.13 \pm 1.30$, then our black hole mass estimate becomes $M_\mathrm{BH} = 8.06^{+2.58}_{-2.67} \times 10^6 ~\msun$. It is evident that true uncertainties on the virial estimate of $M_{\mathrm{BH}}$ are dominated by the systematic uncertainties in the scale factor, which are significantly larger than those derived from the lag and line-width measurements alone.

We note that there are other estimates of the scale factor, such as those obtained by separating galaxies into different populations based on mass \citep{greene2010} and morphology \citep{graham2011}, which yield estimates of \textit{f} different from that of \citet{park2012} by up to a factor of $\sim$2. For example, \citet{woo2013} investigated the scale factor for both quiescent and active galaxies as a combined sample and found $f = 5.9^{+2.1}_{-1.5}$. Furthermore, recent work by \citet{hokim2014} showed that the scale factor can be different for galaxies with pseudobulges and classical buldges, with $f = 3.2 \pm 0.7$ for pseudobulges and $f = 6.3 \pm 1.5$ for classical bulges. Various ongoing efforts that further examine the \msigma relation for local galaxies will improve the precision of the scale factor in the near future as the number of reverberation mapped AGNs increases. Moreover, there has been progress in constraining $f$ for individual galaxies by dynamically modeling the BLR \citep{pancoast2013}.

%-----TABLE-----%

\begin{deluxetable*}{lclcc}
\tablecaption{Line Widths, Lags, and Derived Black Hole Masses\label{tab:mbhtable}}
\tablecolumns{5}
\tablehead{
\colhead{Emission Line} & 
\colhead{$\sigma_{\mathrm{line}}$~(\kms)} & 
\colhead{Lag Computation Method} &
\colhead{$\tau_\mathrm{cen,rest}$ (days)} & 
\colhead{\mbh~($10^6$\, \msun)}
}
\startdata

\phs\hbeta      &  \phs 770 $\pm$ 49   & CCF               &  \lagrest                 &  $8.06^{+1.59}_{-1.72}$  \\
\phs            &  \phs                & \texttt{JAVELIN}  &  \lagjavelinrest          &  $6.85^{+1.00}_{-0.98}$  \\
\phs\hgamma     &  \phs 741 $\pm$ 73   & CCF               &  $10.17^{+2.86}_{-2.56}$     &  $5.59^{+1.92}_{-1.79}$  \\
\phs            &  \phs                & \texttt{JAVELIN}  &  \phn$9.31^{+1.32}_{-1.37}$  &  $3.99^{+0.97}_{-0.98}$  \\
\phs\hdelta     &  \phs 827 $\pm$ 83   & CCF               &  \phn$5.96^{+2.35}_{-2.41}$  &  $4.20^{+1.85}_{-1.89}$  \\
\phs            &  \phs                & \texttt{JAVELIN}  &  \phn$5.39^{+0.98}_{-1.88}$  &  $3.19^{+0.86}_{-1.28}$

\enddata
\tablecomments{Line lags are measured against the \textit{V}-band continuum. \mbh from CCF lags were calculated using $f=5.13$ \citep{park2012}, and \mbh from \texttt{JAVELIN} lags were calculated using $f=4.31$ \citep{grier2013}.}
\end{deluxetable*}
%---------------%

In addition, we obtained \mbh estimates using the broad \hgamma and \hdelta lines. No lag estimate was attempted using \ion{He}{2} since the line has a negative lag. The line widths, rest-frame lags, and derived \mbh values are listed in Table~\ref{tab:mbhtable}. The \hgamma and \hdelta light curves are significantly noisier than that of \hbeta; thus, it is not surprising that, for both CCF and \texttt{JAVELIN} cases, the derived \mbh values have much higher fractional uncertainties compared to the \hbeta \mbh. For both CCF and \texttt{JAVELIN} lags, the \hgamma \mbh estimates, though consistent with the \hbeta \mbh values within $1\sigma$ uncertainties, are slightly smaller than those of \hbeta, and \mbh estimates for \hdelta are smaller still. This may be due to the fact that we are using the same $f$ factor for all the emission lines, while the stratified nature of the BLR may imply different scale factors for each line that depend on the geometry and kinematics of the line-emitting gas.

We would like to compare \kagn to other AGNs having similar black hole masses by studying its location on the \msigma relation as well as the $M_\mathrm{BH} - L_\mathrm{bulge}$ relation (black hole mass vs. host-galaxy bulge luminosity). However, because \kagn appears point-like at ground-based resolution, it is impossible to observe structural properties of the host galaxy without high-resolution images from the \textit{Hubble Space Telescope (HST)\/} or ground-based observations using adaptive optics. Additionally, our Lick spectra cannot be used to measure stellar velocity dispersion in \kagn owing to the galaxy's weak starlight component compared to its AGN luminosity.

%============================================================================================================

\section{Eddington Ratio}

AGNs have been observed to follow a tight correlation between the size of the BLR ($R_\mathrm{BLR}$) and continuum luminosity ($L_\lambda$). The $R_\mathrm{BLR} - L_\lambda$ relation can be written in the form 

\begin{equation}
\log\left(\frac{R_{\mathrm{BLR}}}{1~\text{lt-day}}\right) = K + \alpha \log\left(\frac{\lambda L_\lambda}{10^{44}~\text{erg~s$^{-1}$}}\right),
\end{equation}

\noindent where $L_\mathrm{\lambda}$ is measured at $\lambda_\mathrm{rest} = 5100$~\angstrom. \citet{bentz2013} found the values of $K$ and $\alpha$ to be $1.560^{+0.024}_{-0.024}$ and $0.546^{+0.027}_{-0.027}$ respectively, with a scatter of around 0.13 dex for their best fit. From these parameters and the lag for \kagn, we expect to find $\lambda L_{\lambda}(5100~\mathrm{\angstrom})=1.64^{+0.59}_{-0.65} \times 10^{43} ~\mathrm{erg~s^{-1}}$. We used combined Kast blue- and red-side spectra to measure $L_\lambda$ and adopted the spectral fitting components for the mean spectrum (shown in Figure~\ref{fig:spectraldecomp}) to estimate the starlight contribution in this region, which we found to be approximately 40\% of the total flux. Correcting for Galactic extinction, we roughly estimate $\lambda f_{\lambda}(5100~\mathrm{\angstrom})\approx1.6 \times 10^{-12}~\mathrm{erg~cm^{-2}~s^{-1}}$ for the AGN, corresponding to $\lambda L_{\lambda}(5100~\mathrm{\angstrom})\approx2.4 \times 10^{43} ~\mathrm{erg~s^{-1}}$ for a luminosity distance of 354 Mpc. (We assume the same standard $\Lambda \text{CDM}$ cosmology as \citealt{bentz2013}, where $H_\mathrm{0}=72$ \kms $\mathrm{Mpc^{-1}}$, $\Omega_\mathrm{M} = 0.3$, and $\Omega_\Lambda = 0.7$.) This is consistent with expectations given the scatter in the fit values from \citet{bentz2013}.

\kagn is a NLS1, a class of objects thought to have high $L/L_{\mathrm{Edd}}$ (\citealt{pogge2011}, and references therein). We apply the bolometric correction used by \citet{kaspi2000}, where $L_{\mathrm{bol}}\approx 9 \, \lambda L_{\lambda}\mathrm{(5100~\angstrom)}$, and obtain an estimate of $L_{\mathrm{bol}}=2.2 \times 10^{44}~\mathrm{erg~s^{-1}}$ and $L/L_{\mathrm{Edd}} \approx 0.2$ using \bhmass. We compared this Eddington ratio to those of the four LAMP 2008 NLS1 galaxies, which were calculated using black hole masses published by \citet{bentz2009} and $\lambda L_{\lambda}\mathrm{(5100~\angstrom)}$ values given in \citet{bentz2013}. After applying the same bolometric correction as for \kagn, we found $L/L_\mathrm{Edd} = [0.5, 0.7, 1.2, 0.9]$ for Mrk 1310, Mrk 202, NGC 4253, and NGC 4748, respectively. Compared to these NLS1s, \kagn has a significantly lower Eddington ratio.

A recent study by \citet{du2014} measured the \hbeta lag and \mbh of three NLS1 galaxies (Mrk 335, Mrk 142, and IRAS F12397), all of which appear spectroscopically similar to \kagn. The authors compute the Eddington rate based on a thin accretion disk model \citep{ssdisk}. This rate, denoted by $\dot{m}_\mathrm{ss}$, is written as

\begin{equation}
\dot{m}_\mathrm{ss} \approx 20.1 \left(\frac{L_\mathrm{44}}{\text{cos}~i}\right)^{3/2} M^{-2}_{7} \eta_\mathrm{ss},
\end{equation}

\noindent where they define $L_\mathrm{44} = \lambda L_\mathrm{\lambda}/10^{44}~\text{erg s}^{-1}$ at $\lambda = 5100$~\angstrom, $M_7 = M_{\mathrm{BH}}/10^7 M_{\sun}$, and $\text{cos}~i = 0.75$ as the inclination typical of Type 1 AGNs. For a minimal radiative efficiency of $\eta_{ss} = 0.038$, they find Eddington ratios of 0.6, 2.3, and 4.6 for Mrk 335, Mrk 142, and IRAS F12397, respectively. Applying this same prescription to \kagn, with $M_7 = 0.81$ and $L_{44} = 0.24$, we find $\dot{m}_\mathrm{ss} = 0.2$, in agreement with our $L/L_\mathrm{Edd}$ value obtained using the \citet{kaspi2000} bolometric correction.

%============================================================================================================

\section{Summary}

We photometrically and spectroscopically monitored the \textit{Kepler}-field AGN \kagn over a period of nine months. We found an \hbeta rest-frame lag of \lagrest days with respect to continuum variations using cross-correlation methods, and a lag of \lagjavelinrest days using the \texttt{JAVELIN} method. We also measured emission-line lags with respect to the \textit{Kepler} light curve and found slightly shorter lags compared to those measured against the \textit{V}-band light curve, which is expected given the contributions of broad emission lines and red continuum flux to the \textit{Kepler} band. We measured an \hbeta velocity dispersion of $\sigma_{\mathrm{line}}=$ \hbetawidth, and calculated a black hole virial mass of \bhmass using $\tau_\mathrm{CCF}$ and scale factor empirically derived from local active galaxies by \citet{park2012}, and a black hole mass of \bhmassjavelin using $\tau_\mathrm{\texttt{JAVELIN}}$ and scale factor taken from \citet{grier2013}. For this mass, the Eddington ratio is $L/L_{\mathrm{Edd}} \approx 0.2$.

\kagn was the second AGN for which data was obtained in this interrupt observing mode from Lick Observatory, and the second AGN in the \textit{Kepler} field to be monitored by ground-based telescopes (the first being Zw 229-015, \citealt{barth2011}). Comparing our lag results with those obtained by the LAMP 2008 collaboration \citep{bentz2009}, our lag uncertainties are slightly larger. However, considering the much longer lag of \kagn, our \hbeta fractional lag precision, at less than 20\%, is still very good. Our analysis using \textit{Kepler} light curves also offers one of the first direct comparison of reverberation mapping results between ground- and space-based observations for a \textit{Kepler} AGN. The success of our campaign demonstrates the robust capabilities of interrupt-mode observations for monitoring AGN variability. Factors that negatively impact our measurements, such as inconsistency in data quality and gaps in the spectroscopic light curves (in this case due to the AGN being observed only during dark runs), are mitigated by a well-sampled \textit{V}-band light curve obtained by combining observations from several ground-based telescopes as well as the long duration of the program.

Further observations of \kagn can provide additional insight into various properties of the AGN and host galaxy. Specifically, observations of the bulge properties can put \kagn on the $M_{\mathrm{BH}}-L_{\mathrm{bulge}}$, $M_{\mathrm{BH}}-M_{\mathrm{bulge}}$, and $M_{\mathrm{BH}}-\sigma_*$ relations. Since the host galaxy is very compact, high-resolution HST or adaptive optics imaging will be needed to examine the host-galaxy morphology.

\bigskip
\medskip

We would like to thank the dedicated staff at Lick Observatory for their continuous efforts in supporting our interrupt-mode observations, and Melissa Graham and David Levitan for contributing Lick data from their observing nights. We would also like to thank the following Nickel telescope observers for their data contributions: Kyle Blanchard, Peter Blanchard, Chadwick Casper, Byung Yun Choi, Michael Ellison, Kiera Fuller, Jenifer Gross, Michael Kandrashoff, Daniel Krishnan, Erin Leonard, Gary Li, Michelle Mason, and Andrew Wilkins.

Research by L.P. and A.J.B. at UC Irvine has been supported by NSF grant AST-1108835. Research by M.A.M. at UCLA was supported by NSF grant AST-1107812. A.V.F.'s group at UC Berkeley was supported through NSF grants AST-1108665 and AST-1211916, the TABASGO Foundation, and the Christopher R. Redlich Fund. KAIT and its ongoing operation were made possible by donations from Sun Microsystems, Inc., the Hewlett-Packard Company, AutoScope Corporation, Lick Observatory, the NSF, the University of California, the Sylvia \& Jim Katzman Foundation, and the TABASGO Foundation.  We are very grateful to our late colleague Weidong Li, who was instrumental in making KAIT successful and taught us much about photometry.

J.M.S. is supported by an NSF Astronomy and Astrophysics Postdoctoral Fellowship under award AST-1302771. Research by D.C.L., J.C.H., and
J.M.F. at San Diego State University is supported by NSF grants AST-1009571 and AST-1210311.  The WMO 0.9-m telescope was funded by
NSF grant AST-0618209.

This work makes use of observations from the LCOGT network. Some of the data presented in this paper were obtained from the Mikulski
Archive for Space Telescopes (MAST). STScI is operated by the Association of Universities for Research in Astronomy, Inc., under
NASA contract NAS5-26555. Support for MAST for non-{\it HST} data is provided by the NASA Office of Space Science via grant NNX13AC07G and
by other grants and contracts. This paper includes data collected by the \emph{Kepler} mission. Funding for the \emph{Kepler} mission is
provided by the NASA Science Mission directorate.

\vspace{1cm}

\bibliographystyle{apj}
\bibliography{citations}

\end{document}